\documentclass[aps,twocolumn,showpacs,preprintnumbers,floatfix]{revtex4}
\usepackage{graphicx}
\usepackage{epsfig}
\usepackage{amsmath}
\usepackage{dcolumn}
\usepackage{bm}
\usepackage[
             colorlinks=true,%
             linkcolor=blue,%
             citecolor=blue
             ]{hyperref}

\def\be{\begin{equation}}
\def\ee{\end{equation}}
\def\bea{\begin{eqnarray}}
\def\eea{\end{eqnarray}}
\def\bsp{\be\begin{split}}

\def\ra{\rangle}

\begin{document}
\title{Spectrum and Bethe-Salpeter amplitudes of $\Omega$ baryons from lattice QCD}

\author{\small Jian Liang\footnote{liangjian@ihep.ac.cn},${}^{1}$ Wei Sun,${}^1$ Ying
Chen\footnote{cheny@ihep.ac.cn},${}^{1,2}$ Wei-Feng Chiu,${}^1$, Ming Gong,${}^{1,2}$ Chuan
Liu,${}^3$\\ Yu-Bin Liu,${}^4$ Zhaofeng Liu,${}^{1,2}$ Jian-Ping Ma,${}^5$ and Jian-Bo Zhang
${}^6$}

\affiliation{\small $^1$Institute of High Energy Physics, Chinese Academy of Sciences, Beijing 100049, China\\
 $^2$Theoretical Center for Science Facilities, Chinese Academy of Sciences, Beijing 100049, China\\
 $^3$ School of Physics, Peking University, Beijing 100871, China\\
$^4$ School of Physics, Nankai University, Tianjin 300071, China\\
$^5$ Institute of Theoretical Physics, Chinese Academy of Sciences, Beijing 100080, China\\
$^6$ Department of Physics, Zhejiang University, Hangzhou, Zhejiang 310027, China }

\begin{abstract}
{ The $\Omega$ baryons with $J^P=3/2^\pm, 1/2^\pm$ are studied on the lattice in the quenched
approximation. Their mass levels are ordered as $M_{3/2^+}<M_{3/2^-}\approx M_{1/2^-}<M_{1/2^+}$,
as is expected from the constituent quark model. The mass values are also close to those of the
four $\Omega$ states observed in experiments, respectively. We calculate the Bethe-Salpeter
amplitudes of $\Omega(3/2^+)$ and $\Omega(1/2^+)$ and find there is a radial node for the
$\Omega(1/2^+)$ Bethe-Salpeter amplitude, which may imply that $\Omega(1/2^+)$ is an orbital
excitation of $\Omega$ baryons as a member of the $(D,L_N^P)=(70,0_2^+)$ supermultiplet in the
$SU(6)\bigotimes O(3)$ quark model description. Our results are helpful for identifying the quantum
numbers of experimentally observed $\Omega$ states. }
\end{abstract}

\pacs{11.15Ha, 12.38.Gc, 12.39Mk}
\maketitle

\section{Introduction}
\label{introduction} There are four $\Omega$ baryon states (the strange number $S$=-3) observed
from experiments~\cite{Agashe:2014kda}. Except for the lowest-lying one, $\Omega(1672)$, which is
well known as a member of the $J^P=3/2^+$ baryon decuplet, the $J^P$ quantum numbers of the other
states, namely, $\Omega(2250)$, $\Omega(2380)$, and $\Omega(2470)$, have not been completely
determined from experiments. If they are dominated by the three-quark components, the conventional
$SU(6)\bigotimes O(3)$ quark model with a harmonic oscillator confining potential can be used to
give them a qualitatively description. In this picture, the baryons made up of $u,d,s$ quarks can
be classified into energy bands that have the same number $N$ of the excitation quanta in the
harmonic oscillator potential~\cite{Klempt:2009pi}. Each band consists of a number of
supermultiplets, specified by $(D,L_N^P)$, where $D$ stands for the irreducible representation of
the flavor-spin $SU(6)$ group, $L$ is the total orbital angular momentum, and $P$ is the parity of
the supermultiplet. For $\Omega$ baryons whose flavor wave functions are totally symmetric, the
ground state of $\Omega$ baryons should be in the $(56,0_0^+)$ supermultiplet with the quantum
number $J^P=3/2^+$, namely the $\Omega(1672)$ state. The states in $(70,1_1^-)$ supermultiplet
should have a total spin $S=1/2$ and a unit of the orbital excitation, such that their $J^P$
quantum number can be either $3/2^-$ or $1/2^-$. Therefore $\Omega_{3/2^-}$ and $\Omega_{1/2^-}$
are expected to approximately degenerate in mass up to a small splitting due to the different spin
wave functions. The $J^P=\frac{1}{2}^+$ $\Omega$ baryons should be in either the $(56,2_2^+)$ or
$(70,0_2^+)$ multiplets. Therefore the lowest several $\Omega$ states should have the energy levels
ordered as $M_{3/2^+}<M_{3/2^-}\approx M_{1/2^-}<M_{1/2^+}$. On the other hand, for the
$(56,2_2^+)$ and $(70,0_2^+)$ multiplets, since they belong to the different $SU(6)$
representations, their spatial wave functions can be different and can serve as a criterion to
distinguish them from each other.

However, the quark model is not an ab-initio method and can only give qualitative results, so
studies from first principles are desired, such as lattice QCD method. Early lattice QCD works can
be found in \cite{Chiu:2005zc,Alexandrou:2008tn}. The most recent systematic study with unquenched
configurations was carried out in work \cite{Bulava:2010yg} where the authors find 11 strangeness
-3 states with energies near or below 2.5 GeV using sophisticated smearing schemes for operators
and variational method for the extraction of energy levels, but have difficulties to distinguish
the single $\Omega$ states from possible scattering states. In this work, we explore the excited
states of $\Omega$ baryons in the quenched approximation, whose advantage in this topic is that the
excited states are free from the contamination of scattering states. We focus on the several
lowest-lying $\Omega$ states with $J^P=\frac{1}{2}^\pm, \frac{3}{2}^\pm$. In addition to their
spectrum, we also investigate the Bethe-Salpeter amplitudes of these states through spatially
extended operators, which may shed lights on the internal structure of these $\Omega$ states.


This paper is organized as follows: Sec. \ref{operators and correlation functions} contains our
calculation method including the operator constructions, fermion contractions and wave function
definitions. The numerical results of the spectrum and the wave functions are presented in Sec.
\ref{numerical details and simulation results}. The conclusions and a summary can be found in Sec.
\ref{summary}.

\section{operators and correlation functions}
\label{operators and correlation functions}

\subsection{Interpolating Operators for $\Omega$ Baryons}

The interpolating operator for $\Omega$ baryons can be expressed as
\begin{equation}
\mathcal{O}^\mu=\epsilon^{abc}(s^T_a\mathcal{C}\gamma^\mu s_b)s_c,
\end{equation}
where $\mathcal{C}=\gamma_2\gamma_4$ is the $C$-parity operator, $a,b,c$ are color indices, and
$s^T$ means the transpose of the Dirac spinor of the strange quark field $s$. However,
$\mathcal{O}^\mu$ has no definite spin and can couple to the $J=3/2$ and $J=1/2$
states~\cite{Weinberg:1995mt}. The $J=3/2$ and $J=1/2$ components of $\mathcal{O}_{\Omega}^\mu$ can
be disentangled by introducing the following projectors~\cite{Alexandrou:2008tn}
\begin{eqnarray}
\mathcal{P}_{3/2}^{\mu\nu}&=&\delta^{\mu\nu}-\frac{1}{3}\gamma^\mu\gamma^\nu-\frac{1}{3p^2}(p\!\!\!/\gamma^\mu
p^\nu+p^\mu\gamma^\nu p\!\!\!/),\nonumber\\
\mathcal{P}_{1/2}^{\mu\nu}&=&\delta^{\mu\nu}-\mathcal{P}_{3/2}^{\mu\nu}.
\end{eqnarray}
In the lattice studies, only the spatial components of $\mathcal{O}^\mu$ are implemented. If we
consider the $\Omega$ baryons in their rest frames, the projectors above can be simplified as
\begin{eqnarray}
\mathcal{P}_{3/2}^{ij}&=& \delta^{ij}-\frac{1}{3}\gamma^i\gamma^j,\nonumber\\
\mathcal{P}_{1/2}^{ij}&=& \frac{1}{3}\gamma^i\gamma^j.
\end{eqnarray}
Thus the spin projected operators with the definite spin quantum number can be obtained as
\begin{eqnarray}\label{irreps}
\mathcal{O}^i_{3/2}&=&\sum_j \mathcal{P}_{3/2}^{ij}\mathcal{O}_{\Omega}^j \nonumber\\
\mathcal{O}^i_{1/2}&=&\sum_j \mathcal{P}_{1/2}^{ij}\mathcal{O}_{\Omega}^j.
\end{eqnarray}
Furthermore, one can also use the parity projectors $P^{\pm}=\frac{1}{2}(1\pm\gamma_4)$ to ensure
the definite parities of baryons states.

It should be noted that for now all the operators are considered in the continuum case. On a finite
lattice, the spatial symmetry group $SO(3)$ breaks down to the octahedral point group $O$, whose
irreducible representations corresponding to $J=1/2$ and $J=3/2$ are the two-dimensional $G_1$
representation and the four-dimensional $H$ representation, respectively. Generally, there exist
subduction matrices to project the continuum operators to octahedral point group operators
\cite{Edwards:2011jj}, say,
\begin{equation}
\mathcal{O}(J,\Lambda)_{r}=\sum_{m}S(J,\Lambda)^{m}_{r}\mathcal{O}(J)_{m},
\end{equation}
where $\mathcal{O}(J)_{m}$ is the continuum operator with total spin $J$ and the third component of
spin $m$, $\mathcal{O}(J,\Lambda)_{r}$ is the $r$-th component of the octahedral point group
operator under irreducible representation $\Lambda$, $S(J,\Lambda)^{m}_{r}$ is the subduction
matrices. In our case, $S(\frac{1}{2},G_1)$ and $S(\frac{3}{2},H)$ are both unit matrices, so that
the operators in Eq.~\ref{irreps}, which are actully used in this study, are already the
irreducible representations of the lattice symmetry group $O$.

We also consider the spatially extended interpolation operators by splitting $\mathcal{O}^\mu$ into
two parts with spatial separations. The explicitly expressions are written as
\begin{eqnarray}
\label{operator}
\mathcal{O}_{1}^\mu (r)&=&\sum_{|\vec{r}|}\epsilon^{abc}[s^T_a(x+\vec{r})\mathcal{C}\gamma^\mu s_b(x)]s_c(x),\nonumber\\
\mathcal{O}_{2}^\mu (r)&=&\sum_{|\vec{r}|}\epsilon^{abc}[s^T_a(x)\mathcal{C}\gamma^\mu s_b(x+\vec{r})]s_c(x),\nonumber\\
\mathcal{O}_{3}^\mu (r)&=&\sum_{|\vec{r}|}\epsilon^{abc}[s^T_a(x)\mathcal{C}\gamma^\mu
s_b(x)]s_c(x+\vec{r}).
\end{eqnarray}
where the summations are over $\vec{r}$'s with the same ${r=|\vec{r}|}$ in order to guarantee the
same quantum number as case of $r=0$. These three splitting procedures have been verified to be
numerically equivalent, so we make use of the third type, $\mathcal{O}_3(r)$,  in the practical
study. These operators are obviously gauge variant, so we carry out the lattice calculation by
fixing all the gauge configurations to the Coulomb gauge first.

The general form of the two-point function of a baryon of quantum number $J^P$ with $P=\pm$ is
\begin{equation}\label{two-point}
C_J^{\pm,i}(r,t)=\mathrm{Tr}\left[(1\pm\gamma_4)\sum_{\vec{x},j}\langle\mathcal{P}_J^{ij}\mathcal{O}_3^{j}(r,x)
\bar{\mathcal{O}}_3^{j}(0)P_J^{ji}\ra\right]
\end{equation}
The summation on $\vec{x}$ ensures a zero momentum. For $\Omega$ baryons, there exist six different
Wick contractions as shown in the following figure Fig.~(\ref{contract}).
\begin{figure}[!h]
\includegraphics[height=3.5cm]{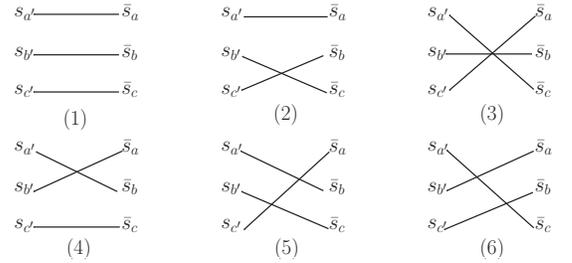}
\caption{Six ways of contraction, we use color indices to label the three $s$ quarks. Solid line
means contraction.} \label{contract}
\end{figure}

\begin{figure}[!h]
\includegraphics[height=5.0cm]{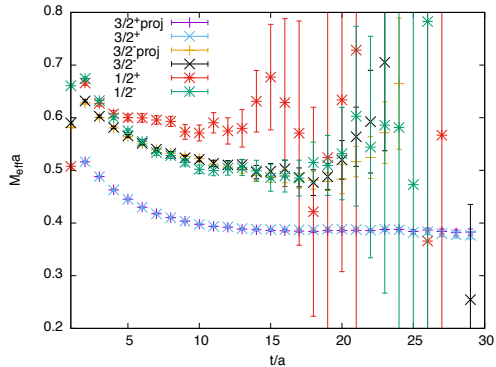}
\caption{The effective mass plateaus of $\Omega$ baryons using point source with/without projections.
Except for $J=\frac{3}{2}$ states, the point-source two point
functions have not good plateaus at the short time range.} \label{point_test}
\end{figure}

\subsection{Source technique}
In principle, all states with the same quantum number $J^P$ contribute to the two-point functions
$C_J^{P,i}(r,t)$. For baryons, it is known that the the signal-to-noise ratio of the two-points
damps very quickly since the noise decreases as $\sim e^{-3/2m_\pi t}$ in $t$, which is much slower
than the decay of the signal $e^{-M_B t}$, where $M_B$ is the baryon mass. Therefore, in order to
obtain clear and reliable signals of the ground state from two-point functions in the available
early time range, some source techniques are implemented by replacing the local operator
$O_3^j(\mathbf{0},0)$ by some versions of spatially extended source operators $O_3^{j,(s)}(0)$
which enhance the contribution of the ground state and suppress that from excited states. The
extended source operator $O_3^{j,(s)}$ is usually realized by calculating the quark propagators
through a source vector with a spatial distribution $\phi(\mathbf{x})$,
\begin{equation}
M(x;y)S_F^{(s)}(y;t_0)=\sum\limits_{\mathbf{z}}\delta(\mathbf{x}-\mathbf{z})\delta(t-t_0)\phi(\mathbf{z}),
\end{equation}
thus the effective propagator $S_F^{(s)}(y;t=0)$ relates to the normal point source propagator
$S_F(y;\mathbf{z},t_0)$ as
\begin{equation}
S_F^{(s)}(y;t_0)=\sum\limits_{\mathbf{z}}\phi(\mathbf{z})S_F(y;\mathbf{z},t_0).
\end{equation}
When one calculates a baryon two-point function using the same Wick contraction by replacing the
point-source propagators with the effective propagators, it is equivalent to using the spatially
extended source operator
\begin{equation}
O^{(s)}(t_0)=\sum\limits_{\mathbf{z,w,v}}\phi(\mathbf{z})\phi(\mathbf{w})\phi(\mathbf{v})
\psi(\mathbf{z},t_0)\psi(\mathbf{w},t_0)\psi(\mathbf{v},t_0),
\end{equation}
where $\psi\psi\psi$ stands for the original baryon operator (the color indices and corresponding
$\gamma$ matrices are omitted for simplicity. Note that gauge links should be considered if one
requires the gauge invariance of spatially extended operators). The matrix element of $O^{(s)}$
between the vacuum and the baryon state $|B\rangle$, which manifests the coupling of this operator
to the state, can be expressed as,
\begin{equation}\label{overlap}
\langle 0 |O^{(s)}|B\rangle =
\sum\limits_{\mathbf{z,w,v}}\phi(\mathbf{z})\phi(\mathbf{w})\phi(\mathbf{v})\Phi_B(\mathbf{z},
\mathbf{w},\mathbf{v})\zeta_B,
\end{equation}
where $\zeta_B$ is the spinor reflecting the spin of $|B\rangle$, and
$\Phi_B(\mathbf{z},\mathbf{w},\mathbf{v})$ is its Bethe-Salpeter amplitude, which is defined as the
corresponding matrix element of the original operator,
\begin{equation}
\langle 0|\psi(\mathbf{z})\psi(\mathbf{w})\psi(\mathbf{v})|B\rangle \equiv
\Phi_B(\mathbf{z},\mathbf{w},\mathbf{v})\zeta_B,
\end{equation}
In order to enhance the coupling $\langle 0 |O^{(s)}|B\rangle$ and suppress the related coupling of
excited states, the essence is to tune the parameters in $\phi(\mathbf{x})$ such that
$\phi(\mathbf{z})\phi(\mathbf{w})\phi(\mathbf{v})$ resembles
$\Phi_B(\mathbf{z},\mathbf{w},\mathbf{v})$ as closely as possible and the overlap integration in
Eq.~(\ref{overlap}) (actually summations over the spatial lattice sites) can be maximized. If the
BS amplitudes can be approximately interpreted to be the spatial wave function of a state, the
coupling of this operator to excited states can be minimized subsequently according to the
orthogonality of the wave functions. The commonly used source techniques include the Gaussian
smeared source~\cite{Gusken:1989qx,Marinari:1988tw}and the wall source in a fixed gauge. The
Gaussian smeared source corresponds to the function $\phi(\mathbf{x})\sim
e^{-{\sigma^2|\mathbf{x}|^2}}$ with $\sigma^2$ a tunable parameter, while the wall source in a
fixed gauge is the extreme situation of the Gaussian smeared source when $\sigma\rightarrow
\infty$. The Gaussian smeared source usually works well for states whose BS amplitude has no radial
nodes. This is similar to the case in the quantum mechanics where a Gaussian-like function serves
as a good trial wave function of the ground state in solving a bound state problem using the
variational method with $\sigma$ the variational parameter.

\begin{figure}[!h]
\includegraphics[height=5.0cm]{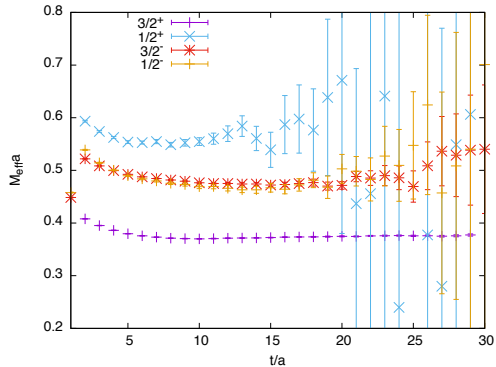}
\caption{Smeared source $\Omega$ spectrum. For $J^p=\frac{3}{2}^{\pm}$ and
$J^p=\frac{1}{2}^-$, we use common Gaussian
smeared source, for $J^p=\frac{1}{2}^+$, we use a novel
``smeared source" with a radial node.}
\label{smear_test}
\end{figure}

For the case of this work, we try first the Gaussian smeared source for $\Omega$ baryons and find
it work surely good for $\Omega_{\frac{3}{2}^+}$. It is not surprising since the
$\Omega_{\frac{3}{2}^+}$ is the ground state whose spatial wave functions is $(1s)(1s)(1s)$ in the
standard quark model with a harmonic oscillator potential. However for other states, especially for
$\Omega_{\frac{1}{2}^+}$, we cannot get a good effective mass plateau before the signals are
undermined by noise. Similar phenomena are also observed by previous works (see
Ref.~\cite{Alexandrou:2008tn} for example). Inspired by the quark model description that the
$J^P=\frac{1}{2}^+$ decuplet baryons belong to the higher excitation energy bands, we conjecture
that the BS amplitude of $\Omega_{\frac{1}{2}^+}$ has radial node(s), and thereby propose a new
type of source which reflects some node structure, say,
\begin{equation}
\phi(\mathbf{x})=(1-A|\mathbf{x}|^2)e^{-\sigma^2 |\mathbf{x}|^2},
\end{equation}
where $\sigma$ and $A$ are parameters to be tuned to give a good effective mass plateau in the
early time range. The effects of the extended source operator on the effective masses of different
states are illustrated in Fig.~(\ref{smear_test}). For $J^P=\frac{3}{2}^\pm, \frac{1}{2}^-$ states,
we use the Gaussian smeared sources which improve the qualities of the effective mass plateaus as
expected. For the $J^P=\frac{1}{2}^+$ state, the new type of the source operators with the nodal
structure makes the effective mass plateaus fairly satisfactory in contrast to the case of point
source. We advocate that this new type of source operators can be potentially applied to other
studies on radial excited states of hadrons.


\section{numerical details and simulation results}
\label{numerical details and simulation results} The gauge configurations used in this work are
generated on two anisotropic ensembles with the tadpole-improved gauge action
\cite{Morningstar:1997ff}. The anisotropy $\xi\equiv a_s/a_t=5$ and the lattice sizes are
$L^3\times T=16^3\times 96$ and $24^3\times 144$, respectively. The relevant input parameters are
listed in Tab.~\ref{latticesetup}, where the $a_s$ values are determined through the static
potential with the scale parameter $r_0^{-1} = 410(20) $MeV. The spatial extensions of the two
lattices are larger than 3 fm, which are expected to be large enough for $\Omega$ baryons such that
the finite volume effects can be neglected. We use the tadpole improved Wilson clover action
\cite{Liu:2001ss} to calculate the quark propagators with the bare strange quark mass parameter
being tuned to reproduce the physical $\phi$ meson mass value. We use a modified version of  a GPU
inverter \cite{Clark:2009wm} to calculate all the inversions in this work.

\begin{table}[]
\caption{ The input parameters for the calculation. Values of the coupling $\beta$, anisotropy $\xi$, the lattice size,
and the number of measurements are listed. $a_s/r_0$ is determined by the static potential,
the first error of $a_s$ is the statistical error and the second one comes from the uncertainty of the scale parameter
$r_0^{-1}=410(20)$ MeV.}
\begin{ruledtabular}
  \begin{tabular}{ccccccc}
     $\beta$ &  $\xi$  & $a_s/r_0$ &$a_s$(fm) & $La_s$(fm)& $L^3\times T$ & $N_{\rm conf}$ \\\hline
       2.4  & 5  & 0.461(4) & 0.222(2)(11) & $\sim 3.55$ &$16^3\times 96$ & 1000 \\
      2.8  & 5  & 0.288(2) & 0.138(1)(7) & $\sim 3.31 $&$24^3\times 144$ & 1000  \\
    \end{tabular}
  \end{ruledtabular}
\label{latticesetup}
\end{table}

\begin{figure}[!h]
\includegraphics[height=5.0cm]{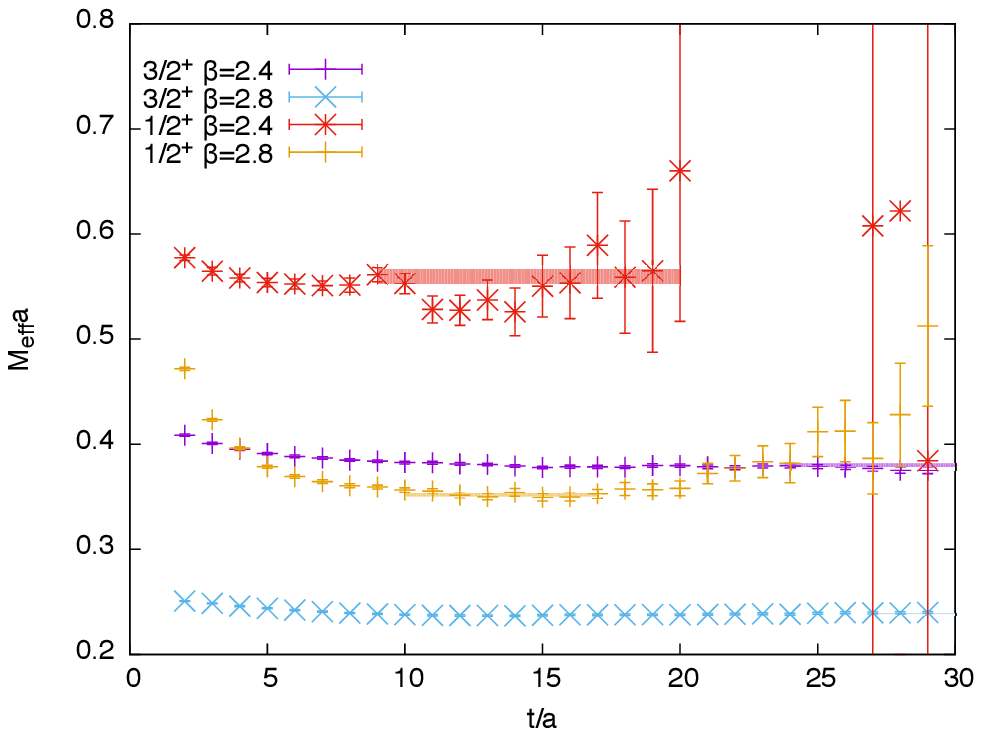}
\includegraphics[height=5.0cm]{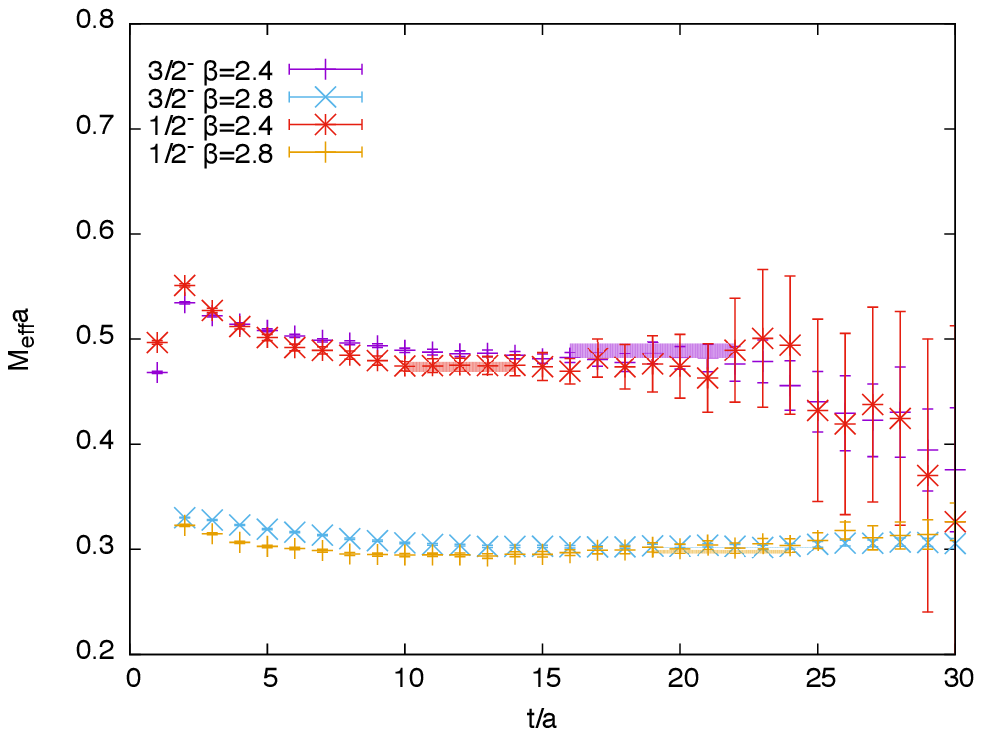}
\caption{Effective mass plots of the $\Omega$ system. The two ensembles are both included.
The points with errorbars are lattice data, while the
colored bands are fit results indicating both the fit range and the fit error.} \label{mass}
\end{figure}

As mentioned before, the spatially extended operators we use for $\Omega$ baryons are not gauge
invariant, so we calculate the corresponding two-point functions in the Coulomb gauge by first
carrying out the gauge fixing to the gauge configurations. By the use of the source vectors with
properly tuned operators, we generate the quark propagators in this gauge, from which the two-point
functions in different channels are obtained. Since we focus on the ground states in each channel,
the related two-point functions are analyzed with the single-exponential function form in properly
chosen time windows,
\begin{equation}
C_2^J(r,t)\overset{t\to\infty}{\sim}N^J\Phi^J(r)e^{-m^Jt},
\end{equation}
where $J$ denotes different quantum numbers, $N^J$ stands for a irrelevant normalization constant,
$\Phi^J(r)$ is the BS amplitude and $m^J$ is the mass. In order to take care of the possible
correlation, we fit $C_2^J(r,t)$ with different $r$ simultaneously through a correlated
miminal-$\chi^2$ fit procedure, where the covariance matrix are calculated by the bootstrap method.
 As such, in addition to the masses $m_J$, we can also obtain the $r$-dependence of the the BS amplitudes
$\Phi^J(r)$. Figure~(\ref{mass}) shows the effective mass plateaus for $C^J(r=0,t)$ and the fit
range. We quote the bootstrap errors as the statistical ones for masses and BS amplitudes.

The masses for different $\Omega$ states on the two lattice are listed in Tab.~(\ref{masstable}),
where the mass values are expressed in the physical units using the lattice spacings in
Tab.~\ref{latticesetup}. The masses of these states are insensitive to the lattice spacings which
implies that the discretization uncertainty is small for these states. It is seen that the mass of
the $J^P=\frac{3}{2}^+$ $\Omega$ we obtain is consistent with the physical mass of $\Omega(1672)$,
the masses of $J^P=\frac{3}{2}^-$ and $\frac{1}{2}^-$ are almost degenerate, as expected from the
quark model, but lower than the experimental states $\Omega(2250)$ and $\Omega(2380)$. For the
$J^P=\frac{1}{2}^+$ state, we get a mass of 2.464(26) GeV on the coarse lattice while 2.492(14) GeV
on the fine lattice, which is in agreement with the mass of $\Omega(2470)$.
\begin{table}
\caption{The spectrum of the $\Omega$ baryons on the two lattices. The errors of masses are all
statistical. We do not include the error owing to the uncertainty of $r_0^{-1}=410(2)$ MeV here.}
\begin{ruledtabular}
\begin{tabular}{ccccc}
     $\beta$ &  $m_{\Omega_{3/2+}}$ &  $m_{\Omega_{3/2-}}$  & $m_{\Omega_{3/2-}}$ & $m_{\Omega_{1/2+}}$  \\
             &    (GeV)             &   (GeV)               &  (GeV)              &  (GeV)               \\\hline
       2.4   &   1.668(9)&  2.176(26) & 2.189(13) & 2.464(26)  \\
       2.8   &   1.695(4)&  2.153(5) &  2.125(14) & 2.492(14)
\end{tabular}
\end{ruledtabular}
\label{masstable}
\end{table}



\begin{figure}[!ht]
\includegraphics[height=5.0cm]{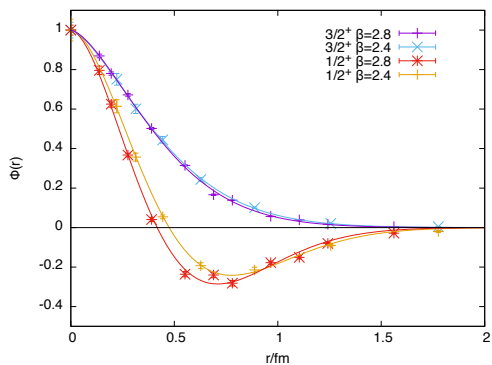}
\caption{The BS amplitudes of $\Omega\frac{3}{2}^+$ and $\Omega\frac{1}{2}^+$. The dots are lattice
results while the lines are the fitting functions in Eq.~(\ref{fitting function}). A radial node of
the BS amplitude of the $\Omega_{\frac{1}{2}^+}$ is observed.} \label{wavef}
\end{figure}

The BS amplitudes for the $\frac{3}{2}^+$ and $\frac{1}{2}^+$ states are plotted in
Fig.~(\ref{wavef}) (normalized as $\Phi_J(r=0)=1$). In order to compare the results from different
lattices, we plot the $x-$axis in physical units. From the figure one can see that the
discretization artifacts are also small for BS amplitudes. We do observe a radial node in the BS
amplitude of $\frac{1}{2}^+$ state. We use the following functions
\begin{eqnarray}\label{fitting function}
 \Phi_{\frac{3}{2}^+}(r)&=&e^{-(r/r_0)^\kappa}, \nonumber\\
\Phi_{\frac{1}{2}^+}(r)&=&(1-b~r^\kappa)e^{-(r/r_0)^\kappa},
\end{eqnarray}
to fit the data points, which are also plotted in curves in the figure.
The fit results are summarized in Tab.~(\ref{wave_fit}).

\begin{table}[!ht]
\caption{Fit results of the BS amplitudes for $\Omega_{\frac{3}{2}^+}$ and
$\Omega_{\frac{1}{2}^+}$.}
\begin{ruledtabular}
\begin{tabular}{ccccc}
     $J^p$ & $\beta$ &  $r_0$ (fm) &  $\kappa$  & $b$   \\
  \hline
       $\frac{3}{2}^{+}$ &2.4  &    0.504(3)&  1.49(2)& \\
       $\frac{3}{2}^{+}$ &2.8  &    0.494(4)&  1.55(2) & \\
       $\frac{1}{2}^{+}$  &2.4  &   0.568(5)&  1.74(3) &  3.7(1) \\
       $\frac{1}{2}^{+}$  &2.8  &   0.529(8)&  1.73(4) &  4.5(2)
\end{tabular}
\end{ruledtabular}
\label{wave_fit}
\end{table}

Now we resort to the non-relativistic quark model to understand the radial behavior of the BS
amplitude of $J^P=1/2^+$ $\Omega$. In the non-relativistic approximation, the relativistic quark
field $\psi$ can be expressed in terms of its non-relativistic components through the
Foldi-Wouthuysen-Tani transformation
\begin{equation}
\psi=\exp \left(\frac{\mathbf{\gamma}\cdot \mathbf{D}}{2m_s}\right) \left(
\begin{array}{c}\chi \\ \eta
\end{array}
\right),
\end{equation}
where the Pauli spinor $\chi$ annihilates a quark and $\eta$ creates an anti-quark, and
$\mathbf{D}$ is the covariant derivative operator. $\eta$ and $\chi$ satisfy the conditions
\begin{equation}
\chi|0\rangle=0,~\langle 0|\chi^{\dagger}=0,~\eta^{\dagger}|0\rangle=0,~\langle 0|\eta=0.
\end{equation}
With this expansion, the operator $\mathcal{O}_{\Omega}^i$ can be expressed as
\begin{eqnarray}
\mathcal{O}_{\Omega}^i &\sim&
\epsilon^{abc}\left[\chi^{aT}\left(1+\frac{(\mathbf{\sigma}\cdot\overleftarrow{{\mathbf
D}})^2}{4m_s^2}\right)\sigma_2\sigma_i\left(1+\frac{(\mathbf{\sigma}\cdot\overrightarrow{{\mathbf
D}})^2}{4m_s^2}\right)\chi^b\right.
\nonumber\\
&&\left. -\chi^{aT}\frac{\mathbf{\sigma}\cdot\overleftarrow{{\mathbf
D}}}{2m_s}\sigma_2\sigma_i\frac{\mathbf{\sigma}\cdot\overrightarrow{{\mathbf
D}}}{2m_s}\chi^b\right] \left(
\begin{array}{c}
\left(1+\frac{(\mathbf{\sigma}\cdot\overrightarrow{{\mathbf
D}})^2}{4m_s^2}\right)\chi^c\\
\frac{\mathbf{\sigma}\cdot\overrightarrow{\mathbf{D}}}{2m_s}\chi^c
\end{array}
\right)\nonumber\\
&+&\ldots.
\end{eqnarray}
We would like to caution that this expansion is not justified rigorously for the strange quark
since its relativistic effect in the hadron might be important. However, the non-relativistic quark
model are usually used to given reasonable descriptions of hadron spectrum, so we tentatively
follow this direction to make the following discussion. The non-relativistic wave function for a
baryon state in its rest frame is defined in principle as
\begin{equation}
\Psi_J(\mathbf{x_1},\mathbf{x_2},\mathbf{x_3})\zeta\sim \langle
0|\epsilon^{abc}\chi^{aT}(\mathbf{x_1})\chi^b(\mathbf{x_2})\chi^c(\mathbf{x_3})|\Omega_J\rangle
\end{equation}
where $\zeta$ stands for the spin wave function for $\Omega_J$. If we introduce the Jacobi's
coordinates,
\begin{eqnarray}
\mathbf{R}&=&\frac{1}{3}(\mathbf{x}_1+\mathbf{x}_2+\mathbf{x}_3)\nonumber\\
\mathbf{\rho}&=&\frac{1}{\sqrt{2}}(\mathbf{x}_1-\mathbf{x}_2)\nonumber\\
\mathbf{\lambda}&=&\frac{1}{\sqrt{6}}(\mathbf{x}_1+\mathbf{x}_2-2\mathbf{x}_3),
\end{eqnarray}
as is usually done in the non-relativistic quark model study of baryons, in the rest frame of
$\Omega_{1/2^+}$ ($\mathbf{R}=0$), the matrix element of
$\mathcal{O}_J^i(\mathbf{x}_1,\mathbf{x}_2,\mathbf{x_3})$ between the vacuum and the $\Omega$ state
can be written qualitatively as
\begin{eqnarray}\label{BS-amplitude}
&&\langle 0|\mathcal{O}_J^i(\mathbf{x}_1,\mathbf{x}_2,\mathbf{x_3})|\Omega_J\rangle\nonumber\\
&\sim&\left(D^i+ A^i\frac{\partial^2}{\partial
\rho^2}+B^i\frac{\partial^2}{\partial\rho\partial\lambda}+C^i\frac{\partial^2}{\partial\lambda^2}\right)
\Psi_J(\mathbf{\rho},\mathbf{\lambda})\zeta\nonumber\\
\end{eqnarray}
where we approximate the covariant derivative $\mathbf{D}$ by the spatial derivative
$\mathbf{\nabla}$.

In the standard non-relativistic quark model with harmonic oscillator potentials for baryons,
baryons can be sorted into energy bands of the the two independent oscillators, the so-called
$\rho$-oscillator and $\lambda$-oscillator, which are depicted by the the radial and orbital
quantum numbers $(n_\lambda, l_\lambda)$ and $(n_\rho, l_\rho)$~\cite{Klempt:2009pi}. For baryons
made up of $u,d,s$ quarks, these energy bands are labelled as $(D,L_N^P)$, where $D$ is the
irreducible representation of the flavor-spin $SU(6)$ group, $L=|l_\rho-l_\lambda|,
|l_\rho-l_\lambda|+1,\ldots, l_\rho+l_\lambda$ is the total orbital angular momentum,
$N=2(n_\rho+n_\lambda)+(l_\rho+l_\lambda)$ is the total number of the excited quanta of the
harmonic oscillators, and $P$ is the parity of baryons. For the flavor symmetric $\Omega$ baryons,
the lowest $J^P=\frac{1}{2}^+$ states can be found in the supermultiplets $(56, 2_2^+)$, and
$(70,0_2^+)$. $(56,2_2^+)$ has the excitation mode $(n_\lambda,n_\rho)=(0,0)$ and
$(l_\lambda,l_\rho)=(2,0)$ or $(0,2)$ with the total spin $S=\frac{3}{2}$, and gives the quantum
number $J^P=\frac{1}{2}^+, \frac{3}{2}^+, \frac{5}{2}^+, \frac{7}{2}^+$. $(70,0_2^+)$ has the
excitation mode $(n_\lambda,n_\rho)=(0,0)$ and $(l_\lambda,l_\rho)=(1,1)$ with $S=\frac{1}{2}$,
which corresponds to the quantum number $J^P=\frac{1}{2}^+$. In this picture, the spatial wave
function of the $(56,2_2^+)$ multiplet can be written qualitatively (here we ignore the angular
part)~\cite{Karl:1969iz, Faiman:1968js}
\begin{equation}\label{wavef1}
\Psi(\rho,\lambda)\sim (\rho^2+\lambda^2)e^{-\alpha (\rho^2+\lambda^2)},
\end{equation}
while the spatial wave function of the $(70,0_2^+)$ multiplet is either
\begin{equation}\label{wavef2}
\Psi(\rho,\lambda)\sim (\rho^2-\lambda^2)e^{-\alpha(\rho^2+\lambda^2)},
\end{equation}
or
\begin{equation}\label{wavef3}
 \Psi(\rho,\lambda)\sim
\rho\lambda e^{-\alpha(\rho^2+\lambda^2)},
\end{equation}
where the parameter $\alpha$ depends on the constituent quark mass and the parameters in the
potential. Obviously, the local operators correspond to $\lambda=\rho=0$, such that their coupling
to the $J^P=\frac{1}{2}^+$ state can be largely suppressed. Recalling that the interpolation
operator we use for $\Omega$ baryons is $\mathcal{O}_3(r)$, which corresponds to $\rho=0$ and
$\lambda\propto r$. As such, we have the qualitatively radial behavoirs of the Bethe-Salpeter
amplitudes
\begin{equation}
\langle 0|\mathcal{O}_{3,\frac{1}{2}^+}^i(r)|\Omega_{\frac{1}{2}^+}\rangle \sim(
A'+B'r^2+C'r^4)e^{-\alpha r^2}\zeta^i,
\end{equation}
if we use the wave functions in Eq.~(\ref{wavef1}) and Eq.~(\ref{wavef2}), and
\begin{equation}
0|\mathcal{O}_{3,\frac{1}{2}^+}^i(r)|\Omega_{\frac{1}{2}^+};(70,0_2^+)\rangle \sim(
A''+B''r^2)e^{-\alpha r^2}\zeta^i
\end{equation}
for the wave function form in Eq.~(\ref{wavef3}).  Obviously, the former may has two nodes in the
$r$ direction, while the later has only one. In this sense, the radial behaviors of the BS
amplitudes in Fig.~\ref{wavef} may imply that the $J^P=\frac{1}{2}^+$ $\Omega$ baryon we have
observed is possibly mainly the $(70,0_2^+)$ state, whose spatial wave function may have the
qualitative form in Eq.~(\ref{wavef3}). It should be noted that these discussions are very
tentatively and the the reality can be much more complicated. This can be seen in
Table~\ref{wave_fit} where the parameters $\kappa$ deviate substantially from $\kappa=2$ which
corresponds to the harmonic oscillator potential.

\section{summary}
\label{summary} We carry out a lattice study of the spectrum and the Bethe-Salpeter amplitudes of
$\Omega$ baryons in the quenched approximation. In the Coulomb gauge, we propose a new type of
source vectors for the calculation of quark propagators, which is similar in spirit to the
conventionally used Gaussian smearing source technique, but is oriented to increase the coupling to
the states whose Bethe-Salpeter amplitude may have more complicated nodal behavior than that of the
ground state. As for a excited states, either an orbital excitation or a radial excitations, it is
expected that their BS amplititude may have radial nodes, so we use source vectors with nodal
structures, which resemble the node structure of its BS amplitude. This technique works in
practice, since we can obtain fairly good effective mass plateaus for $J^P=\frac{1}{2}^+$ at the
early time slices.

With the quark mass parameter tuned to be at the strange quark mass using the physical mass of the
$\phi$ meson, we calculate the spectrum of $\Omega$ baryons with the quantum number
$J^P=\frac{3}{2}^\pm, \frac{1}{2}^\pm$ on two anisotropic lattices with the spatial lattice spacing
set at $a_s=0.222(2)$ fm and $a_s=0.138(1)$ fm, respectively. On both lattices, the
$J^P=\frac{3}{2}^-$ and $\frac{1}{2}^-$ $\Omega$ baryons have almost degenerate mass in the range
from 2100 MeV to 2200 MeV. This is compatible with the expectation of the non-relativistic quark
model that they are in the same supermultiplet $(70,1_1^-)$ with the same excitation mode, say,
$(n_\lambda,n_\rho)=(0,0)$ and $(l_\rho,l_\lambda)=(1,0)$ or (0,1), and the same total quark spin
$S=\frac{1}{2}$. For the $\frac{1}{2}^+$ $\Omega$ baryon, we obtain its mass at roughly 2400-2500
MeV. Furthermore, we also calculate the BS amplitude of the $\frac{1}{2}^+$ $\Omega$ baryon in the
Coulomb gauge and observe a radial node, which can be qualitatively understood as the reflection of
the second order differential of the non-relativistic wave function of $(70,0_2^+)$ baryons.
Therefore it is preferable to assign the $\frac{1}{2}^+$ $\Omega$ state we observe to be a member
of $(70,0_2^+)$ supermultiplet instead of that of $(56,2_2^+)$.

We notice that the latest $N_f=2+1$ full-QCD lattice calculation has obtained 11 energy levels of
the $\Omega$ spectrum around and below 2500 MeV, but has difficulties in the assignment of their
status for the sake of no reliable criterion to distinguish single particle states from the
would-be scattering states. Fortunately we are free of this kind of trouble with the quenched
approximation, such that the masses we obtain can be taken as those of the {\it bare} $\Omega$
baryon states before their hadronic decays are switch on. In comparison with the experiments, our
predicted masses of $J^P=\frac{3}{2}^-$ and $\frac{1}{2}^-$ $\Omega$ baryons are close to that of
$\Omega(2250)$, and the mass of $J^P=\frac{1}{2}^+$ is consistent with $\Omega(2470)$. This
observation may be helpful in determining their $J^P$ quantum numbers.

\section*{ACKNOWLEDGEMENTS}
The numerical calculations are carried out on Tianhe-1A at the National Supercomputer Center (NSCC)
in Tianjin. This work is supported in part by the National Science Foundation of China (NSFC) under
Grants No. 11105153, No. 11335001, and 11405053. Z.L. is partially supported by the Youth
Innovation Promotion Association of CAS. Y.C. and Z.L. also acknowledge the support of NSFC under
No. 11261130311 (CRC 110 by DFG and NSFC).



\end{document}